\documentclass[aps,prd,onecolumn,superscriptaddress,subscriptemail,preprintnumbers,amsmath,amssymb]{revtex4}

\usepackage{graphicx}
\usepackage{dcolumn}
\usepackage{bm}

\def\lapproxeq{\lower .7ex\hbox{$\;\stackrel{\textstyle
<}{\sim}\;$}}
\def\gapproxeq{\lower .7ex\hbox{$\;\stackrel{\textstyle
>}{\sim}\;$}}

\begin{document}

\hyphenation{Es-ta-du-al}
\title{A Study on anisotropy of cosmic ray distribution with a small
array of water cherenkov detectors}


 \author{F. Sheidaei}
\author{M. Bahmanabadi}\email{bahmanabadi@sina.sharif.edu}

\author{A. Keivani  }

   \author{M. Khakian Ghomi}
   \author{J. Samimi}
\author{A. Shadkam}
   \affiliation{Department of Physics, Sharif University of Technology,
P.O.Box 11365-9161, Tehran, Iran}


\begin{abstract}
$\cos6{\cdot7± 1{\cdot}4}$ The study of the anisotropy of the
arrival directions is an essential tool to investigate the origin
and propagation of cosmic rays primaries. A simple way of
recording many cosmic rays is to record coincidences between a
number of detectors. We have monitored multi-TeV cosmic rays by a
small array of water cherenkov detectors in Tehran($35^{\circ}
43^{\prime}$ N, $51^{\circ} 20^{\prime}$E, $1200$m a.s.l). More
than $1.1\times10^{6}$ extensive air shower events were recorded.
In addition to the Compton- Getting effect due to the motion of
the earth in the Galaxy, an anisotropy has been observed which is
due to a unidirectional anisotropy of cosmic ray flow along the
Galactic arms.
\end{abstract}



\maketitle


\section{Introduction}
Although Cosmic Rays( CRs) have been known for almost one century,
their origin remains uncertain, mostly because their trajectories
are bent by Galactic magnetic fields and they do not individually
point back to their sources. Cosmic rays in the lower energy range
have gyro radii of about 1pc or less in typical galactic magnetic
fields( a proton with an energy of $10^{15}$eV would have a gyro
radius of 1 pc in a $1\mu$G field). Moreover, since these fields
are chaotic on scales ranging at least from $10^{8}$cm to
$10^{20}$cm \cite{Armstrong}, the transport of CRs is diffusive up
to high energies, which tends to make their angular distribution
isotropic. Therefore, even collectively, the CR arrival directions
hold virtually no information about the source distribution in
space. However, as the energy of the CRs increases, it acan appear
either because the diffusive approximation does not hold anymore,
or because the diffusion coefficient becomes large enough to
reveal intrinsic inhomogeneities in the source distribution.
Specifically, even if the diffusive regime holds, the density of
CR sources in the Galaxy is believed to be larger in the inner
regions than in the outer ones, and this can cause a slightly
higher CR flux coming from the Galactic center than from the
anti-center. Meanwhile, the global CR streaming away from the
Galactic plane( towards the halo) can be a source of measurable
anisotropy. However, the detailed angular distribution of CRs is
quite hard to predict, even if we assume a definite source
distribution, because it also depends on the propagation
conditions, which are related to both large scale and small scale
magnetic field configurations, and on the position of the Earth
relative to major magnetic structures, such as the local Galactic
arm.

From a general point of view, the characterization of the CR
anisotropy provides useful information to constrain the GCR
diffusion models, notably the effective diffusion coefficients,
related to the magnetic field structure. Indeed the level of CR
anisotropy depends on the diffusion coefficient, $D$: in a simple
model where CR sources are homogeneously distributed in a disk of
thickness $2h$ and the CRs are confined in a halo of height $H$,
the anisotropy at a distance $z$ above the Galactic plane($z<h$)
is estimated as $\Delta\simeq 3D/cH \times z/h$ \cite{Ptuskin}.
The deviation from isotropy is typically below $1\%$ and can be as
low as $0.03\%$ \cite{Smith}. Anisotropy measurements at various
energies can thus provide crucial information about the energy
dependence of the diffusion coefficient. This information is
particularly important to constrain the GCR source spectrum, since
it sets the relation between the source power law index and the
observed one, through the energy dependent confinement of CRs in
the Galaxy. This diffusion maybe is broadly along the magnetic
field lines which are in tubes of dimensions greater than the gyro
radii. So the direction of the peak of the anisotropy would
indicate the direction back towards the cosmic ray source, and the
amplitude of the anisotropy would give information on the
scattering process involved in the diffusion. Specifically, an
estimate of the mean free path might be obtained.

The anisotropy is due to a combination of effects. Compton and
Getting\cite{Compton} proposed in 1935 that the motion of the
solar system relative to the rest frame of the cosmic ray plasma
should cause an energy independent dipole anisotropy with maximum
in the direction of motion. The earth's rapid motion in space,
resulting from the rotation of our galaxy, results variations in
cosmic ray intensity fore and aft of the earth's motion. Following
Compton - Getting (1935), the magnitude of the anisotropy is
expressed as
\begin{equation}
\eta=(\gamma+2)\frac{u}{c}\cos\beta,\\
\end{equation}
where $\gamma$ denotes the power law index of the energy spectrum
of cosmic rays, $u$ the velocity of the detector relative to the
production frame of the cosmic rays ( where they are presumed to
be isotropic), $c$ the speed of light, and $\beta$ the cosmic ray
direction relative to $u$, i.e. $\cos\beta$ is the projection of
the cosmic ray along the forward direction of $u$. In fact the
value of $(\gamma+2)\frac{u}{c}$ is
 $(f_{max}-f_{min})/(f_{max}+f_{min})$
 with $f_{max}$ the counting rate along the direction
of the velocity and $f_{min}$ along the contrary direction. The
magnitude of the anisotropy is extremely small and independent of
the cosmic ray energy. Our data will be analyzed in a sun-centered
frame, and so if data accumulation is done over an integer number
of solar years, it is only necessary that the orbital speed of the
Earth around the sun($\sim 30$km s$^{-1}$) be considered. The
large effect due to the Galactic rotational speed (220 km
s$^{-1}$) will cancel out as the data are averaged over this time
\cite{Poirier}. Many experiments have been carried out for
detection of this effect\cite{Poirier,Lin}.

 Doppler effect studies of globular
clusters and extra galactic nebulae have revealed a motion of the
earth of about 220km s$^{-1}$ towards right ascension
$\alpha\simeq21$h and declination $\delta\simeq47^{\circ}$N due
chiefly to the rotation of the Galaxy. This motion, with a speed
of about 0.1\% $c$ will affect the intensity of the incoming
cosmic rays by changing both the energy of the cosmic ray
particles and the number received per second. Using value of 220
km$ s^{-1}$ for $u$, and 2.7 for the spectral index, Eqn. (1)
gives a Compton Getting Effect(CGE) amplitude of
$0.345\times10^{-2}$ for the fractional forward-backward asymmetry
caused by the motion of the earth in the Galaxy.

Another effect that can produce sidereal modulation is solar
diurnal and seasonal changes
 in the atmospheric temperature and pressure. As the atmospheric temperature and pressure
 change during the course of a day, the balance of cosmic ray secondary particle interaction
 and decay changes. This propagates to changes in the detection rate that depend on the
detector type( air shower, underground muon, surface muon) and on
the energy threshold.
 These changes tend to have a strong Fourier component with a frequency of one solar
day($ \simeq365$ cycles/ year) and one solar year( 1 cycle/ year).
In some( but by no means all)
 experiments, the interplay between the daily and seasonal modulation can produce
significant modulation in sideband frequencies of $\simeq( 365\pm
1)$cycles/ year\cite{Farley}. The modulation with frequency 366
cycles/ year appears as a sidereal modulation. The size of the
atmospheric contribution to apparent sidereal anisotropy can be
estimated from the amplitude of the pseudo- sidereal( 365 cycles/
year) frequency. If it is large, the atmospheric effect can be
subtracted using the amplitude and phase of the pseudo- sidereal
component.
 The anisotropy that remains after accounting for the Compton- Getting and atmospheric
effects is due to solar and galactic effects. At the lowest
energies($\sim100$GeV),
 the Interplanetary Magnetic Field( IMF) produced by the solar wind effects the
 sidereal anisotropy: when the local IMF points toward the sun, the anisotropy peaks
 at about 18hr right ascension, while it peaks at 6hr when it points away\cite{Kojima1, Kojima2}.
 The average over the two configurations produces a small, residual anisotropy
 peaking at around 2-4hr. At higher energies, local IMF plays a negligible role.
 Instead, the heliosphere extending to distances of order 100AU is believed to
 induce anisotropy in cosmic rays with energies around 1TeV\cite{Nagashima1,Nagashima2}.
  Beyond this energy, the anisotropy is believed to be primarily of galactic
origin. For instance, the galactic magnetic field around the solar
system neighborhood could
 produce anisotropy. Also, an uneven distribution of sources of cosmic rays
( presumably, mostly supernova remnants) may produce anisotropy.
It is believed that star formation( and thus, supernova remnants)
occur primarily in the spiral arms of the galaxy. The earth is
located at the inner edge of the Orion spur. Thus,
 in the direction of the Orion spur( galactic longitude between $60^{\circ}$ to $270^{\circ}$) they are distributed
 nearby sources of cosmic rays, while in the complementary direction, they are much further
away.

 Because of small anisotropy, large data sets are required
to make useful measurements which overcome the statistical
uncertainties of counting experiments. A simple way of recording
many cosmic rays is to record coincidences between a number of
detectors. Few statistically significant anisotropies have been
reported from extensive air shower experiments in the two last
decades. Aglietta et al.( 1996,EAS-TOP)\cite{Aglietta} published
an amplitude of $(3.7\pm0.6)\times10^{-4}$ and phase $\phi=(
1.8\pm0.5)$hr LST, at $E_{\circ}\approx200$TeV. Analysing the
Akeno experiment, Kifune et al.(1986)\cite{Kifune} reported
results of about $2\times10^{-3}$ at about 5 to 10 PeV. An
overview of experimental results can be found in \cite{Clay}.We
have operated a small array of water cherenkov detectors on the
roof of physics department at Sharif University of Technology in
Tehran($35^{\circ} 43^{\prime}$ N, $51^{\circ} 20^{\prime}$E,
$1200$m a.s.l=890 gcm$^{-2}$) as a prototype for constructing an
Extensive Air Shower(EAS) array on Alborz mountain range at
altitude of over 2500m near Tehran.

The main purpose of this article is to study the unidirectional
anisotropy of cosmic rays flow along the Galactic arms which was
observed in the sidereal time at energies above 50 TeV. We
describe the experimental setup in section II, and the data
analysis and discussion in section III.

\section{Experimental setup}
Four Water Cherenkov Detectors( WCDs) used for recording Extensive
Air Showers( EASs). The four WCDs consist of cylindrical tanks
made of polyethylene with a diameter of 64cm and a height of 130cm
filled up to a height of 120cm with 382 liters of purified water.
All the inner surfaces of the four Cherenkov tanks were optically
sealed and covered with white paint which reflects light in a
diffusive way. Each one of them have a single 5.2cm PMT( model EMI
9813KB) located at the top of the water level along the cylinder
axis. The array arranged in a square with side 608cm as shown in
Fig.1, on the roof of the Physics Department at Sharif university
of Technology in Tehran($35^{\circ} 43^{\prime}$ N, $51^{\circ}
20^{\prime}$E, $1200$m a.s.l=890 gcm$^{-2}$). The signal produced
by secondary particles of an EAS are triggered with an amplitude
threshold of -500mV by an 8-fold fast discriminator( CAEN N413A).
The threshold of each discriminator is set at the separation point
between the signal and background noise levels. The discriminator
outputs are connected to one of three Time to Amplitude
Converters( TAC) (EG\&G ORTEC 566) which are set to a full scale
of 200ns ( maximum acceptable time difference between two WCDs).
The output of due to the No.3 WCD is connected to the start inputs
of TAC1, TAC2, and TAC3. The outputs of due to No.1, 2, and 4 WCDs
are respectively connected to the stop inputs of TAC1, TAC2, and
TAC3. then the outputs of these three TACs are fed into a
multi-parameter Multi channel Analyzer( MCA) ( KIAN AFROUZ Inc.)
via an Analogue to Digital Converter ( ADC)( KIAN AFROUZ Inc.)
unit. The output of TAC1 triggers the ADC, and the 3 time lags
between the output signals of PMTs (3,1), (3,2), and (3,4) are
read out as parameters 1 to 3. So by this procedure an event is
logged.

\section{Data Analysis and Discussion}
\subsection{Array event rate}
The data set covers a total period of $2.7\times10^{7}$ seconds. A
total of $1.1\times10^{6}$ events with zenith angles
$\leq60^{\circ}$ were collected during this time, giving a mean
event rate of one event every 24.5 seconds. Figure 2 shows the
events time- spacing distribution. Since events arrive randomly in
time, it is expected that this will follow  an exponential
distribution, $viz.$
\begin{equation}
f(t)=f(0) \exp(-t/\tau).\\
\end{equation}
The event rate can be obtained by fitting this function on the
events time-spacing distribution. One event per every $\tau=24.1$s
is obtained from the fit. A non-random component for the cosmic
ray flux for example a point source of gamma ray gives rise to
deviation from the exponential law. Our observed distribution is
in good agreement with the exponential law.
\subsection{Atmospheric effects on counting rate}
The rate of shower detections depends on a number of factors. If
either temperature or pressure variations have Fourier components
in solar or sidereal times, spurious components may be introduced
into the shower detection rate \cite{Farley}. Various methods are
used in order to study the dependence of event rate on atmospheric
ground pressure, $p$, and temperature, $T$, \cite{Antoni}. In a
multiple regression analysis of event rate against pressure and
temperature, the temperature effect is found to be statistically
insignificant. The CR intensity dependence on barometric pressures
in half-hour intervals are shown in Fig. 3. We can describe the
dependence by the following function
\begin{equation}
R=R_{0} \exp(\frac{p_{0}- p_{i}}{p_{1}})\\
\end{equation}
The values of $R_{0}=74$ events per half an hour, $p_{0}$=883.7mb,
and $p_{1}$=171.5mb were obtained from the data. $p_{i}$ denotes
the measured air pressure at a given time. By this empirical
function, we weighted the raw event rates for atmospheric
pressure. Figure 4 shows the mean half-hour event rate
distributions with and without correction for atmospheric ground
pressure. The distribution of corrected rates is consistent with a
Gaussian distribution, as expected for the statistical fluctuation
of the events rate, and there is no residual temperature effect.

\subsection{Zenithal angle distribution of the EAS events}
Since the thickness of the atmosphere increases with increasing
zenith angle, $\theta$, the number of EAS events is strongly
related to $\theta$, as shown in Fig. 5. The differential zenith
angle distribution can be represented by
\begin{equation}
dN= constant.( \delta_{1}\cos\theta+ \delta_{2}\sin\theta) \cos^{n}\theta \sin\theta d\theta.\\
\end{equation}
Where we split into particles entering through the lid of
cylinderical tank of WCD or through its walls. The first term in
parenthesis of Eqn.(4) is related to the lid and the second to the
walls. The parameter $\delta_{1}$ includes the area of lid
surface, $S_{1}$, and detection probability of particles entering
through the lid, $P_{1}$. The parameter $\delta_{2}$ also includes
the greatest surface area of vertical profile of the WCD, $S_{2}$,
and detection probability of particles entering through its walls,
$P_{2}$. So we can split $\delta_{j}( j=1, 2)$ in the form of

\begin{equation}
\delta_{j}=S_{j}P_{j},
\end{equation}
where only $P_{j}$s are determined from the simulation ( see the
Appendix). $S_{1}$ and $S_{2}$ are respectively
$3.2\times10^{3}$cm$^{2}$ and $7.68\times10^{3}$cm$^{2}$. By
fitting Eqn (4) on our experimental data, $n=7.3$ is obtained.

\subsection{Energy Threshold of our experiment}
Since we can not determine the energy of the showers on an
event-by-event basis, we estimate the energy threshold of our
array by the CORSIKA code for simulation of EAS events
\cite{Heck}. In order to record a shower it is necessary at least
one particle passes through each of the four WCDs. Because our
array has been arranged in a square with side 608cm we can detect
a shower if density of secondary particles to be at least
$\rho_{r}=1/S_{eff}$ particles cm$^{-2}$ at
$r=608\times\sqrt{2}/2=420$cm, where $S_{eff}$ is the effective
surface area of a WCD. This area is calculated as follows:
\begin{equation}
S_{eff}=\frac{\int_0^{\pi/3}(P_1 S_1\cos\theta +P_2
S_2\sin\theta)\sin\theta d\theta}{\int_0^{\pi/3}\sin\theta
d\theta}=6.5\times10^{3}~{\rm cm}^2,
\end{equation}

with $P_1=0.88$, $S_1=3.2\times10^{3}$cm$^{2}$, $P_2=0.93$, and
$S_2=7.68\times10^{3}$cm$^{2}$. The upper limit, $\pi/3$, is due
to events selection with zenith angles $\leq\pi/3$. We simulated
more than $10^{5}$ EAS events with CORSIKA code using the hadronic
interaction models QGSJET and GHEISHA. The energy range for
primary particles was selected from 5TeV to 5PeV, with
differential flux given by $dN/dE\propto E^{-2.7}$ . These
simulations are in different directions with zenith angle from
$0^{\circ}$ to $60^{\circ}$ and azimuth angle from $0^{\circ}$ to
$360^{\circ}$. Finally from the CORSIKA simulations with
$\rho_{r}= 1.5\times10^{-4}$particles cm$^{-2}$ at $r=420$cm, we
obtained the energy threshold $E_{th}=50$TeV.

\subsection{Sidereal time distribution}
After atmospheric correction we calculated the sidereal time (ST)
from $ST=ST_{\circ}+\alpha(ZT-ZT_{0}$). $ST_{0}$ can be looked up
in an almanac \cite{http} for the time $ZT_{0}$, $ZT$ is the solar
time, and $\alpha=1.00273790935$. Figure 6 shows percentage
variation in intensity of the cosmic rays with sidereal time. The
data have been fit to Eqn.(7) which describes a curve with first
and second harmonics( i.e with a once-per-day and a twice-per-day
variation),
\begin{equation}
y=A_{I}\cos[\frac{2\pi}{24}(t-T_{I})]+
A_{II}\cos[\frac{2\pi}{12}(t-T_{II})].
\end{equation}
Where t is in hours. The fitting results of data are summarized in
Table I. The CGE would contribute to the component $A_{I}$ in the
sidereal time asymmetry. This analysis shows that the anisotropy
has a peak close to the sidereal time 21h, when the zenith is
toward the earth's motion. The amplitude of the first harmonic is
$0.32\%$. So there is a definite sidereal time variation whose
phase and amplitude are close to those predicted. In order to
calculate the magnitude of anisotropy due to CGE i.e the value
$\eta$ in Eqn.(1), a mean value for $\cos\beta$ is needed. Assume
$\delta$ is the declination of the direction of earth's motion,
$\lambda$ the latitude of the observer and $H$ the hour angle
between the observer's meridian and the direction of motion, then
to consider Fig.7 the angle $\phi$ between the observer's zenith
and the direction of earth's motion is given by
\begin{equation}
\cos\phi=\sin\delta\sin\lambda+\cos\delta\cos\lambda\cos H.
\end{equation}
On the other hand, $\cos\beta$ is calculated by
\begin{equation}
\cos\beta=\cos\phi\cos\theta+\sin\phi\sin\theta\cos\alpha.
\end{equation}
 Where $\theta$ is the zenith angle of cosmic ray and
$\alpha$ difference between the azimuth angle of the direction of
motion and of cosmic ray( Fig.7), that is $\alpha=A_{1}-A_{2}$,
which $A_{1}$ and $A_{2}$ are obtained by
\begin{equation}
\sin\delta=\sin\lambda\cos\phi+\cos\lambda\sin\phi\cos A_{1},
\end{equation}
and
\begin{equation}
\sin\delta'=\sin\lambda\cos\theta+\cos\lambda\sin\theta\cos A_{2}.
\end{equation}
Where $\delta'$ is the declination of cosmic ray. According to
Eqns (8) to (11), the 24-hour mean of the component of cosmic ray
in the direction of motion ($\cos\beta$) may be obtained. Using
$\lambda=35^{\circ} 43^{'}$ and $\delta=47^{\circ}$, we calculated
the 24-hour mean value of $\cos\phi\simeq 0.43$ with Eqn.(8). With
distribution of $(P_1 S_1 \cos\theta+ P_2 S_2
\sin\theta)\cos^{7.3}\theta \sin\theta$ which describes the
acceptance of detectors and the cosmic ray absorption in the
Earth's atmosphere because of inclination from the vertical
direction in Tehran ( section III, C) we calculated the mean value
of $\cos\theta$=0.88. Also the mean values of $A_{1}$ and $A_{2}$
were obtained by using the mean value $\delta'$. Figure 8 shows
the distribution of cosmic rays declination. The mean value of
declination is $\delta'=32.5^{\circ}$. From Eqns. (10) and (11),
$A_1$ and $A_2$ obtain $49^{\circ}$ and $86^{\circ}$ respectively.
Finally from Eqn.(9) a value of 0.72 is obtained for $\cos\beta$
and this is multiplied by the expected CGE amplitude of $0.345\%$
to yield a predicted effect of expected value of 0.248\%. The
value obtained from experimental data is $0.32\%$ which about
$0.07\%$ is more than the CGE value. This remaining asymmetry of
$0.07\%$, presumably has an origin different to the that of the
CGE.

Since the recorded data are in Tehran with latitude
$35^{\circ}43'N$, the majority of cosmic rays are from the spiral
arm inwards direction, which is at about 20h in right ascension
and $35^{\circ}$ in declination\cite{Jacklyn}. So the remaining
asymmetry is probably due to unidirectional anisotropy of cosmic
ray flow along the Galactic arms. A simple diffusion model
\cite{Allan} suggests that value of this asymmetry, $0.07\%$,
would be roughly equal to the ratio of the scattering mean
 free path to a characteristic dimension of the containment region
( i.e the central Galactic region, with a scale of 10kpc). So
 with amplitude of the anisotropy of $0.07\%$ found in this work, we obtain
 a mean free path of about 7pc which is about perhaps 7 gyro radii.

Since the anisotropies are low, it is necessary to consider the
effect of counting statistics for a finite measured data set. If
we have N events then the probability of getting a fractional
amplitude greater than $r$ is given by \cite{Linsley},
\begin{equation}
P( >r)= e^{-k_{0}},       k_{0}=r^{2}N/4.
\end{equation}
So a convenient parameter for characterizing the anisotropy
amplitude probability distribution is $k_{0}$. We can take
$\sqrt{2}r_{rms}$ which corresponds to $k_{0}$ =1, as noise
amplitude. For the number of events that we have accumulated,
$1.1\times10^{6}$, the total amplitude of $0.32\%$ obtained in
this work can be arisen by chance with a probability of $\sim0.06$
corresponding to $k_{0}=2.8$. This shows a significant anisotropy(
$k_{0}>1$) at the sidereal period. So we conclude that this data
set gives evidence of anisotropy.

\section{Conclusion}
   Cosmic ray data in Alborz observatory clearly shows an anisotropy in
sidereal time with the energy threshold of $\sim50$TeV and the
mean energy of $\sim 121$TeV. One part of this anisotropy is due
to Earth's motion around the Galaxy (the CGE), but our measured
asymmetry suggests the possible existence of some other additional
effects, probably a unidirectional anisotropy of cosmic ray flow
along the Galactic arms. The first harmonic amplitude of our total
measured anisotropy is about $0.32\%$. The CGE contribution to
this
 anisotropy is about $0.248\%$ and the rest, $0.07\%$, is predicted
to be due to the flow along the Galactic arm. The latter
anisotropy suggests a mean free path of about 7pc for these
high-energy cosmic rays. The evidence of these anisotropies is
based on the value of the parameter $k_{0}$, as suggested by
Linsley( 1975) and found in this work to be 2.8, that is,  more
than $k_{0}=1$, the value for the noise amplitude.

 The EAS-TOP experiment reported somewhat lower limits in the energy range
below 1200TeV\cite{Aglietta1}. The relatively large amplitudes
published by the Akeno experiment\cite{Kifune} and our experiment
are difficult to reconcile with the results of the EAS-TOP
experiment.

\section*{Appendix}
 In order to obtain the $P_{j}$s, Eqn.(5), we first calculated the track length of particle
 passing through lid and walls of WCD. To determine the track length distribution we have made
  the following assumption:

A) The zenithal and azimuth angles of particles are uniformly
distributed.

 B) The random distribution increases linearly with $r$,
the distance from center of lid, i.e. it is proportional to the
annulus surface $2\pi r \Delta r$.

The geometry can be simply solved. It is split into particles
entering through the lid or through the walls as seen in Fig. 1.
In these figures,
 $r$ is the distance from the center of the cylinder, the tank radius is $R_0= 32$cm,
the tank height $H_0=120$cm, and $\varphi$ and $\theta$ are
azimuth and zenith angles of particle, respectively. The
simulation process starts by randomly choosing $r$ which increases
linearly with $r$. Then $\varphi$ and $\theta$ are randomly chosen
and the
 track lengths within the tank are evaluated by a simple calculation
( the particle could either leave through the wall or the bottom
lid, as shown in Fig. 1).

Then the number of photons produced along a flight path was
estimated. Charged particles emit light under a characteristic
angle when passing through a medium if their velocity exceeds the
speed of light in the medium. The Cherenkov angle is related to
the particle velocity and the refractive index of the medium $n$(
$\cos\theta=1/n\beta$). For relativistic particles, $\beta=1$, and
the refractive index of purified water, $n=1.33$( for short
wavelengths of visible region), the Cherenkov light is emitted
under $41^{\circ}$. The number of photons produced along a flight
path $dx$ in a wave length bin $d\lambda$ for a particle carrying
unit charge is :

\begin{equation}
\frac{d^{2}N}{dxd\lambda}=\frac{2\pi\alpha\sin^{2}\theta}{\lambda^{2}},
\end{equation}
where $\alpha=1/137$. At wavelengths of 310-470nm the efficiency
of the photomultiplier is maximal. Within 1cm flight path 220
photons are emitted in this wavelength bin. To consider the
effective area of the photomultiplier( 21.2cm$^{2}$), and
neglecting absorption and scattering effects in water we obtained
the number of photons received by PMT ( Fig. 9). Finally with a
$\eta=25\%$ quantum efficiency and a G=$10^{8}$ gain for the PMT,
the number of electrons produced in PMT($ N_{e}=N_{photon}\eta G$)
was calculated. As we know, the output signal at the PMT's anode
is a current or charge pulse. Now to consider the amplitude
threshold of discriminator ( -500mV), and the anode load resistor
and capacitance, we obtained that for producing a pulse with
amplitude -500mV, the number of photons received to PMT should be
more than 1. Because the quantum efficiency of PMT is 25\% using
Fig. 9 we calculated the detection probability, $P_{j}$s, are
respectively 0.88 and 0.93 for lid and walls WCD.

\section*{Acknowledgments} 
This research has been partly supported by Grant No. NRCI 1853 of
National Research Council of Islamic Republic of Iran.

\newpage
\begin{table}[h]
\begin{center}{\footnotesize
\begin{tabular}{cc}
\hline \hline
Amplitude(\%)&Phase(h)\\
\hline
$A_{I}=0.32\pm 0.10$&$T_{I}=21.3 \pm 1.0 $ \\
$A_{II}=0.56\pm 0.10$&$T_{II}=20.6\pm 0.7$   \\
\hline

\end{tabular}
\label{turns} \caption{Parameters of the fit coefficients in
equation(7)}}
\end{center}
 \label{turns}
\end{table}



\newpage
\begin{figure}[!htb]
\begin{center}
\includegraphics[width=0.8\textwidth]{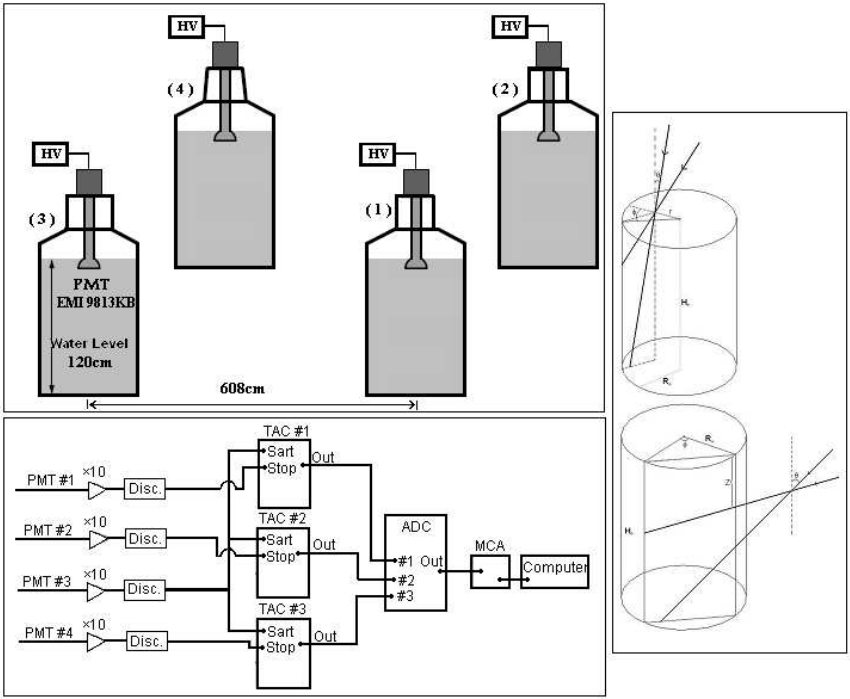}
\end{center}
 \caption{Schematic view of water Cherenkov detectors as a square array, and the electronic circuit.
 Particle tracks through lid (right hand-top), and tracks through walls( right hand-bottom) have also been shown.}
 \end{figure}

\newpage
\begin{figure}[!htb]
\begin{center}
\includegraphics[width=0.8\textwidth]{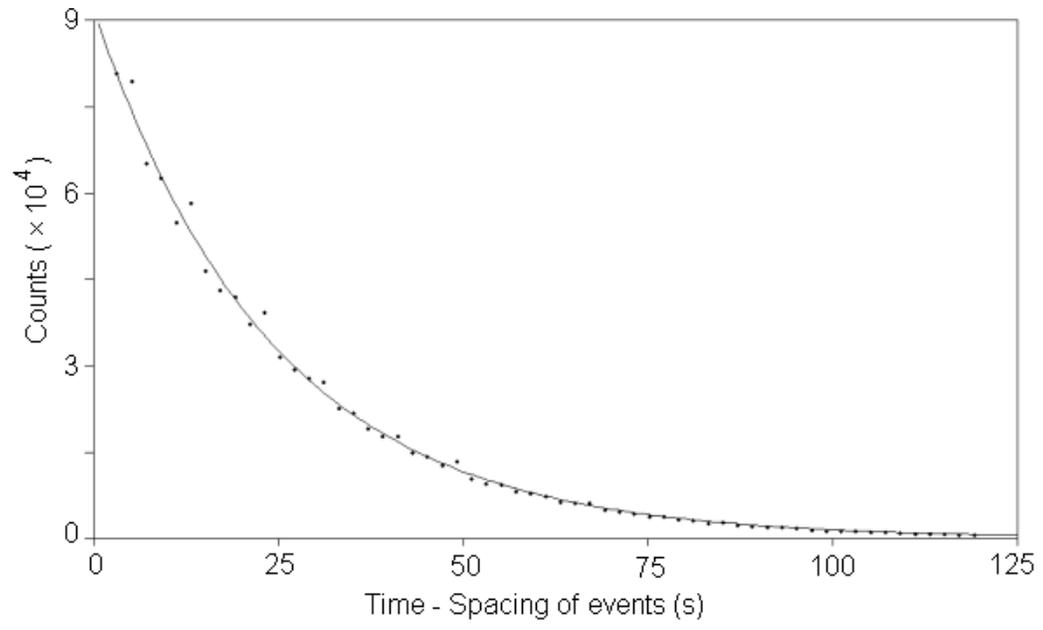}
\end{center}

   \caption {Distribution of events time-spacing.}
              \label{Gam}%
\end{figure}

\newpage
\begin{figure}[!htb]
\begin{center}
\includegraphics[width=0.8\textwidth]{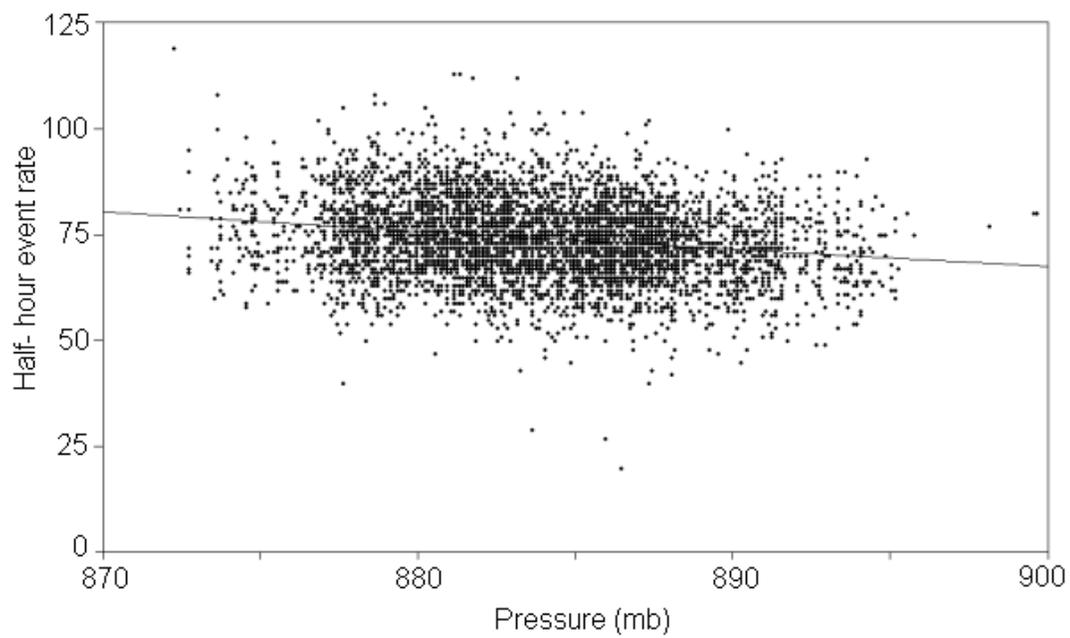}
\end{center}

   \caption {Event rates per half an hour as a function of atmospheric pressure.}
              \label{Gam}%
\end{figure}
\newpage
\begin{figure}[!htb]
\begin{center}
\includegraphics[width=0.6\textwidth]{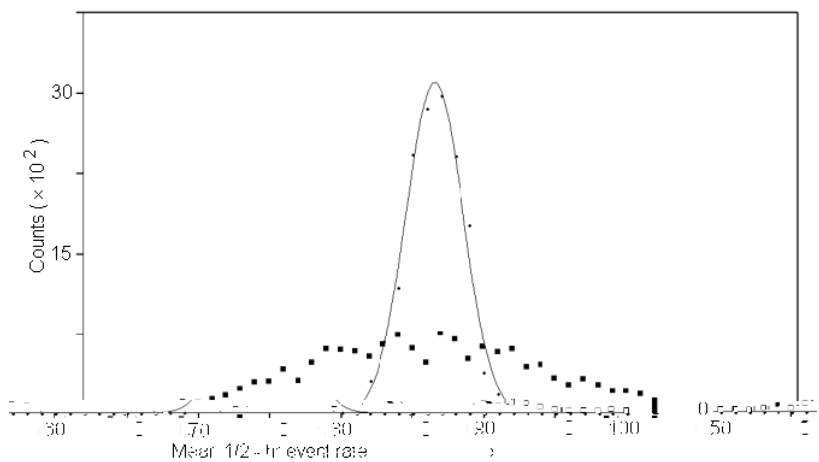}
\end{center}

   \caption {Distribution of mean 1/2-hr event rates before (squares) and after (points) the atmospheric correction for pressure. The
   line curve shows a fit by a Gaussian function.}
              \label{Gam}%
\end{figure}
\newpage
\begin{figure}[!htb]
\begin{center}
\includegraphics[width=0.6\textwidth]{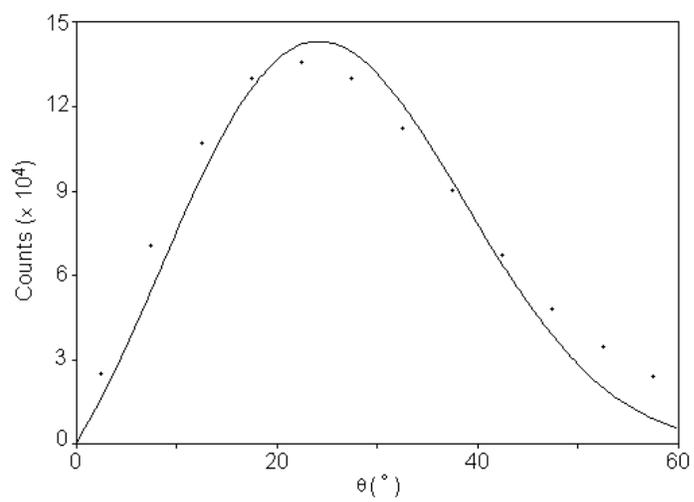}
\end{center}

   \caption {Frequency of air showers vs. zenith angle.}
              \label{Gam}%
\end{figure}
\newpage
\begin{figure}[!htb]
\begin{center}
\includegraphics[width=0.7\textwidth]{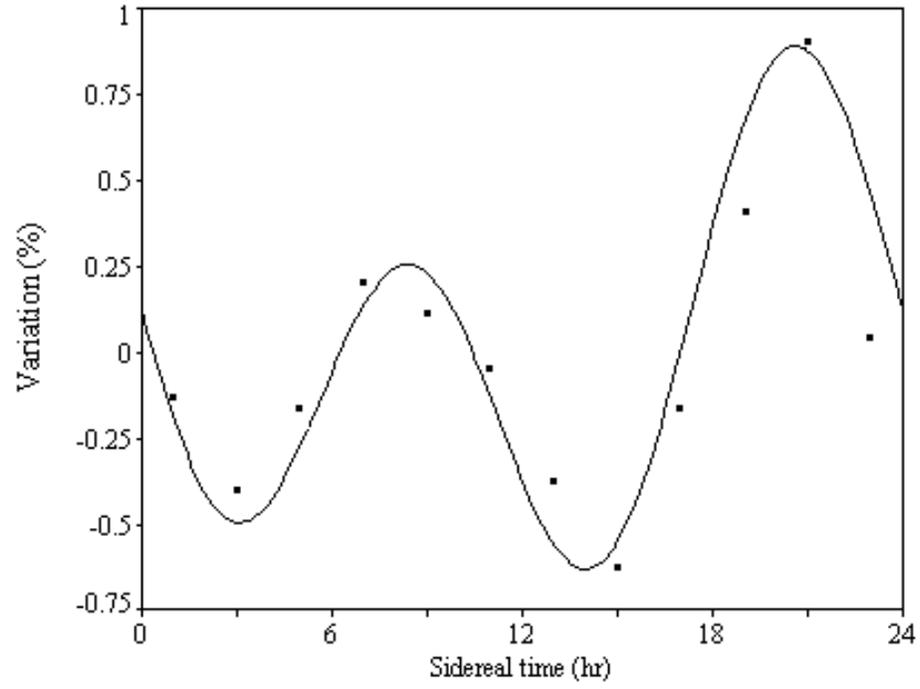}
\end{center}

   \caption {Observed sidereal time variation in intensity of the cosmic rays (points). The curve is the best
   fit to Eqn. (7) with the coefficients as listed in Table I.}
              \label{Gam}%
\end{figure}
\newpage
\begin{figure}[!htb]
\begin{center}
\includegraphics[width=0.6\textwidth]{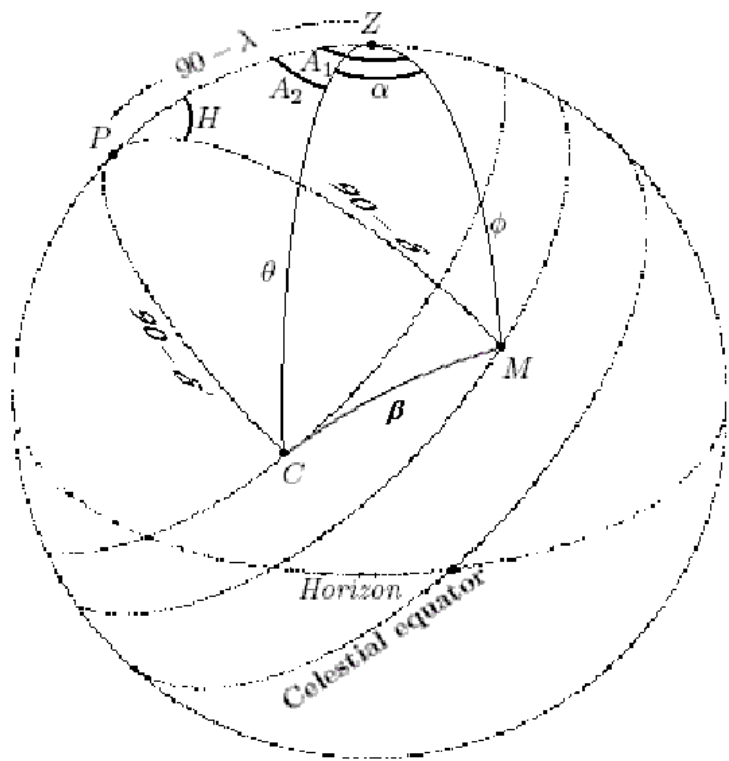}
\end{center}

   \caption {Celestial coordinate, C= direction of cosmic ray, M= direction of earth's
motion, Z= zenith, P= direction of North pole.}
              \label{Gam}%
\end{figure}
\newpage
\begin{figure}[!htb]
\begin{center}
\includegraphics[width=0.6\textwidth]{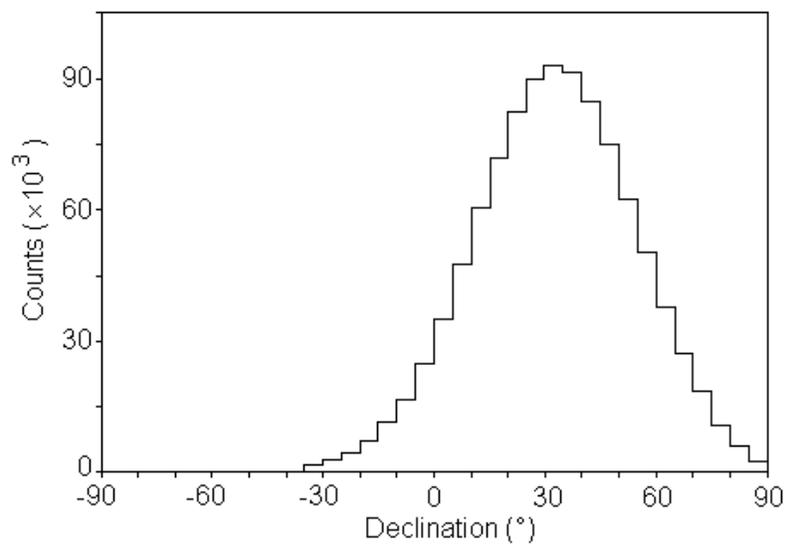}
\end{center}

   \caption {Distribution of air shower events vs. declination angle.}
              \label{Gam}%
\end{figure}

\newpage
\begin{figure}[!htb]
\begin{center}
\includegraphics[width=0.8\textwidth]{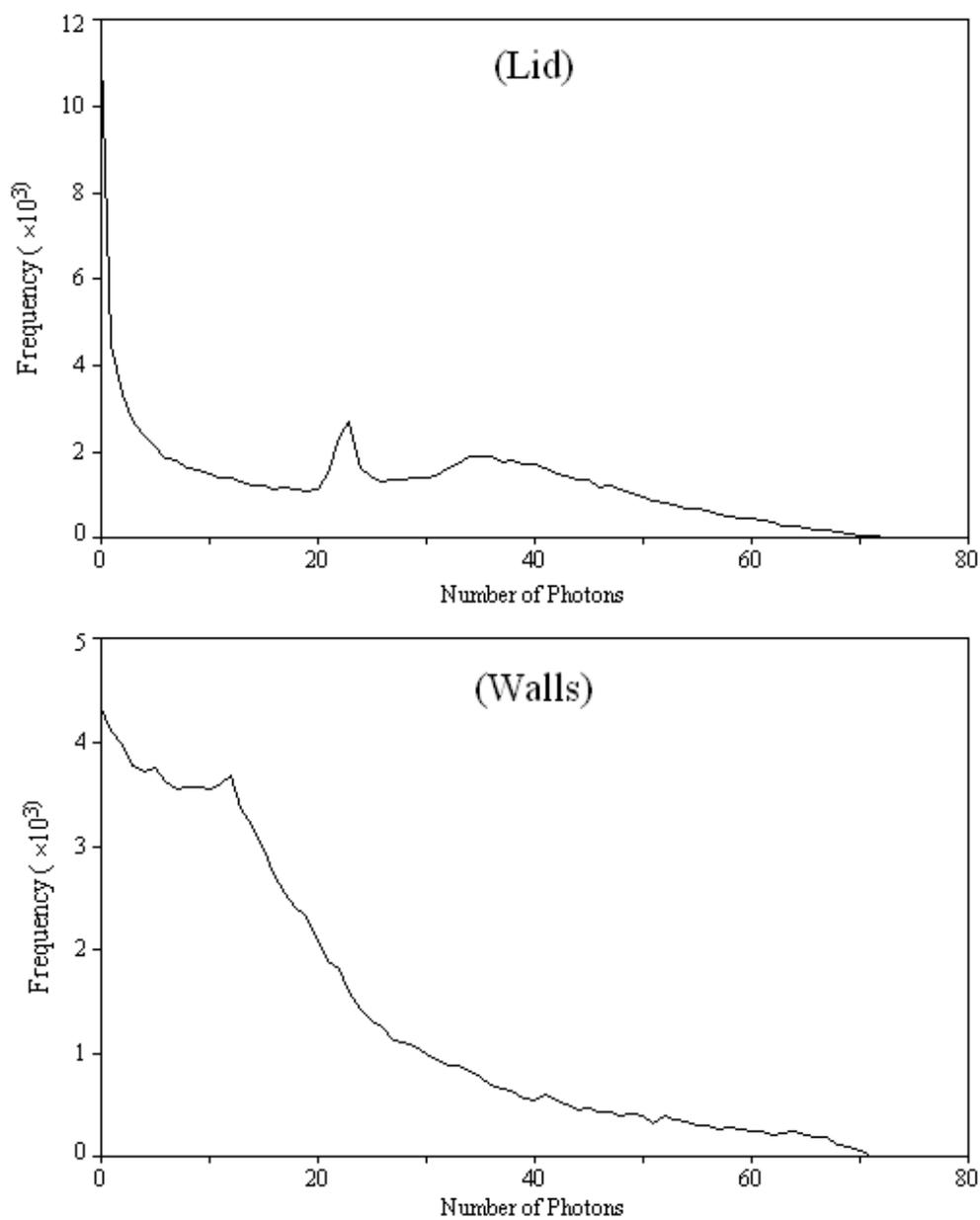}
\end{center}

   \caption {The number of photons received by PMT  in consequence of particle passing along different tracks
   entering through the lid( Top), and the walls( bottom). }
              \label{Gam}%
\end{figure}


\begin{thebibliography}{00}
\expandafter\ifx\csname
natexlab\endcsname\relax\def\natexlab#1{#1}\fi
\expandafter\ifx\csname bibnamefont\endcsname\relax
  \def\bibnamefont#1{#1}\fi
\expandafter\ifx\csname bibfnamefont\endcsname\relax
  \def\bibfnamefont#1{#1}\fi
\expandafter\ifx\csname citenamefont\endcsname\relax
  \def\citenamefont#1{#1}\fi
\expandafter\ifx\csname url\endcsname\relax
  \def\url#1{\texttt{#1}}\fi
\expandafter\ifx\csname
urlprefix\endcsname\relax\def\urlprefix{URL }\fi
\providecommand{\bibinfo}[2]{#2}
\providecommand{\eprint}[2][]{\url{#2}}

\bibitem[{\citenamefont{Armstrong et~al.}(1995)}]{Armstrong}
  \bibinfo{author}{\bibfnamefont{J.W.}~\bibnamefont{Armstrong}},
\bibinfo{author}{\bibfnamefont{B.J.}~\bibnamefont{Rickett}},
\bibinfo{author}{\bibfnamefont{and S.R.}~\bibnamefont{Spangler}},
 \bibinfo{Jurnal}{{ApJ}}
  \textbf{\bibinfo{volume}{443}}, \bibinfo{pages}{209}
(\bibinfo{year}{1995})


\bibitem[{\citenamefont{Ptuskin}(1997)}]{Ptuskin}
  \bibinfo{author}{\bibfnamefont{V.~S.}~\bibnamefont{Ptuskin}},
  \bibinfo{Journal}{{Adv. Space Res.}}
  \textbf{\bibinfo{volume}{19}}, \bibinfo{pages}{697}
(\bibinfo{year}{1997})


\bibitem[{\citenamefont{Smith \& Clay}(1997)}]{Smith}
\bibinfo{author}{\bibfnamefont{A.G.K.}~\bibnamefont{Smith}} \bibnamefont{and}
  \bibinfo{author}{\bibfnamefont{R.W.}~\bibnamefont{Clay}},
 \bibinfo{Journal}{{Aust. J. Phys.}}
 \textbf{\bibinfo{volume}{50}}, \bibinfo{pages}{827}
(\bibinfo{year}{1997})


\bibitem[{\citenamefont{Compton \& Getting 1935}(1935)}]{Compton}
  \bibinfo{author}{\bibfnamefont{A.H.}~\bibnamefont{Compton}}, \bibnamefont{and}
\bibinfo{author}{\bibfnamefont{I.A.}~\bibnamefont{Getting}},
  \bibinfo{Journal}{{Phys. Rev.}}
  \textbf{\bibinfo{volume}{47}}, \bibinfo{pages}{817}
(\bibinfo{year}{1935})


\bibitem[{\citenamefont{Poirier et al.}(2001)}]{Poirier}
  \bibinfo{author}{\bibfnamefont{J.}~\bibnamefont{Poirier}}, \bibnamefont{and}
\bibinfo{author}{\bibfnamefont{C.}~\bibnamefont{D'Andrea}},
\bibinfo{author}{\bibfnamefont{M.}~\bibnamefont{Dunford}},
 \bibinfo{Journal}{{Proc. 27th ICRC,
Hamburg, Germany}}
  , \bibinfo{pages}{3930} (\bibinfo{year}{2001})


\bibitem[{\citenamefont{Lin et al.}(1999)}]{Lin}
  \bibinfo{author}{\bibfnamefont{T.F.}~\bibnamefont{Lin et al.}},
  \bibinfo{note}{{Proc. 26th ICRC, Salt Lake city, USA}}
\textbf{\bibinfo{}}, \bibinfo{pages}{HE2.2.09,
 100} (\bibinfo{year}{1999})

\bibitem[{\citenamefont{Farley \& Storey}(1954)}]{Farley}
  \bibinfo{author}{\bibfnamefont{F. J. M}~\bibnamefont{Farley}}, \bibnamefont{and}
\bibinfo{author}{\bibfnamefont{J. R.}~\bibnamefont{Storey}},
  \bibinfo{note}{{Proc. phys. Soc.}}
   \textbf{\bibinfo{A67}}, \bibinfo{pages}{996}
(\bibinfo{year}{1954})

\bibitem[{\citenamefont{Kojima et al.}(2001)}]{Kojima1}
  \bibinfo{author}{\bibfnamefont{H.}~\bibnamefont{Kojima et al.}},
  \bibinfo{note}{{Proc. 27th ICRC, Hamburg, Germany}}
   \textbf{\bibinfo{10}}, \bibinfo{pages}{3943}
(\bibinfo{year}{2001})

\bibitem[{\citenamefont{Kojima et al.}(2003)}]{Kojima2}
  \bibinfo{author}{\bibfnamefont{H.}~\bibnamefont{Kojima et al.}},
  \bibinfo{note}{{Proc. 28th ICRC, Tsukuba, Japan}}
   \textbf{\bibinfo{7}}, \bibinfo{pages}{3957}
(\bibinfo{year}{2003})

\bibitem[{\citenamefont{Nagashima et al.}(1989)}]{Nagashima1}
  \bibinfo{author}{\bibfnamefont{K.}~\bibnamefont{Nagashima et al.}},
  \bibinfo{note}{{Nuovo Cimento Soc. Ital. Fis.}}
   \textbf{\bibinfo{C 12c}}, \bibinfo{pages}{695}
(\bibinfo{year}{1989})

\bibitem[{\citenamefont{Nagashima et al.}(1989)}]{Nagashima2}
  \bibinfo{author}{\bibfnamefont{K.}~\bibnamefont{Nagashima et al.}},
\bibinfo{author}{\bibfnamefont{K.}~\bibnamefont{Fujimoto}}, \bibnamefont{and}
\bibinfo{author}{\bibfnamefont{R.M.}~\bibnamefont{Jacklyn}},
  \bibinfo{note}{{J. Geophys. Res.}}
   \textbf{\bibinfo{103 A8}}, \bibinfo{pages}{17429}
(\bibinfo{year}{1998})

\bibitem[{\citenamefont{Aglietta et al.}(1996)}]{Aglietta}
  \bibinfo{author}{\bibfnamefont{M.}~\bibnamefont{Aglietta et al.}},
  \bibinfo{note}{{APJ}}
   \textbf{\bibinfo{470}}, \bibinfo{pages}{501}
(\bibinfo{year}{1996})

\bibitem[{\citenamefont{Kifune et al. }(1986)}]{Kifune}
  \bibinfo{author}{\bibfnamefont{T.}~\bibnamefont{Kifune et~al.}},
  \bibinfo{note}{{J. Phys. G}}
\textbf{\bibinfo{12}}, \bibinfo{pages}{129}
(\bibinfo{year}{1986})


\bibitem[{\citenamefont{Clay et al. }(1997)}]{Clay}
\bibinfo{author}{\bibfnamefont{R.}~\bibnamefont{Clay et al.}},
  \bibinfo{note}{{Proc. of 25th ICRC, Durban }}
  \textbf{\bibinfo{4}}, \bibinfo{pages}{185}
  (\bibinfo{year}{1997})



\bibitem[{\citenamefont{Antoni et al.}(2004)}]{Antoni}
  \bibinfo{author}{\bibfnamefont{T.}~\bibnamefont{Antoni et~al.}},
  \bibinfo{note}{{ApJ}}
\textbf{\bibinfo{604}}, \bibinfo{pages}{687}
(\bibinfo{year}{2004})


\bibitem[{\citenamefont{Heck et~al.}(1998)}]{Heck}
  \bibinfo{author}{\bibfnamefont{D.}~\bibnamefont{Heck et~al.}},
  \bibinfo{Journal}{{Report FZKA6019 (Forschungszentrum Karlsruhe)}}
\textbf{\bibinfo{}}, \bibinfo{pages}{} (\bibinfo{year}{1998}) .


 \bibitem[{\citenamefont{ }()}]{http}
\bibinfo{author}{\bibfnamefont{}~\bibnamefont{}}
\bibinfo{note}{{tycho.usno.navy.mil/sidereal.html}}

\bibitem[{\citenamefont{Jacklyn}(1986)}]{Jacklyn}
  \bibinfo{author}{\bibfnamefont{R.M.}~\bibnamefont{Jacklyn}},
  \bibinfo{Journal}{{PASA}}
\textbf{\bibinfo{6}}, \bibinfo{pages}{425} (\bibinfo{year}{1986})

\bibitem[{\citenamefont{Allan}(1972)}]{Allan}
  \bibinfo{author}{\bibfnamefont{H.R.}~\bibnamefont{Allan}},
  \bibinfo{Journal}{{ApJ. Lett.}}
\textbf{\bibinfo{12}}, \bibinfo{pages}{237} (\bibinfo{year}{1986})

\bibitem[{\citenamefont{Linsley}(1975)}]{Linsley}
  \bibinfo{author}{\bibfnamefont{J.}~\bibnamefont{Linsley}},
  \bibinfo{Journal}{{Phys. Rev. Letts.}}
\textbf{\bibinfo{34}}, \bibinfo{pages}{1530}
(\bibinfo{year}{1975})

\bibitem[{\citenamefont{Aglietta et al.}(2003)}]{Aglietta1}
  \bibinfo{author}{\bibfnamefont{M.}~\bibnamefont{Aglietta et al.}},
  \bibinfo{note}{{Proc. of
28th ICRC, Tsukuba, Japan}}
   \textbf{\bibinfo{4}}, \bibinfo{pages}{183}
(\bibinfo{year}{2003})

\end{thebibliography}
\end{document}